\documentclass{article}

\usepackage{arxiv}

\usepackage[utf8]{inputenc} % allow utf-8 input
\usepackage[T1]{fontenc}    % use 8-bit T1 fonts
\usepackage{url}            % simple URL typesetting
\usepackage{booktabs, longtable}       % professional-quality tables
\usepackage{amsfonts}       % blackboard math symbols
\usepackage{nicefrac}       % compact symbols for 1/2, etc.
\usepackage{microtype}      % microtypography
\usepackage{graphicx}
\usepackage{doi}

\title{Proactive and Reactive Engagement of Artificial Intelligence Methods for Education: A Review}

\author{ 
        Sruti Mallik\thanks{Both authors have equally contributed to this study} \\
	    Washington University in St. Louis \\
        St. Louis, MO, USA\\
        \texttt{sruti.mallik@wustl.edu}\\
	\And
	   Ahana Gangopadhyay\thanks{Both authors have equally contributed to this study} \\
       Washington University in St. Louis \\
       St. Louis, MO, USA\\
	\texttt{ahana@wustl.edu} \\
}

\begin{document}
\maketitle

\begin{abstract}
	Quality education, one of the seventeen sustainable development goals (SDGs) identified by the United Nations General Assembly, stands to benefit enormously from the adoption of artificial intelligence (AI) driven tools and technologies. The concurrent boom of necessary infrastructure, digitized data and general social awareness has propelled massive research and development efforts in the artificial intelligence for education (AIEd) sector. In this review article, we investigate how artificial intelligence, machine learning and deep learning methods are being utilized to support students, educators and administrative staff. We do this through the lens of a novel categorization approach. We consider the involvement of AI-driven methods in the education process in its entirety - from students admissions, course scheduling etc. in the \textit{proactive} planning phase to knowledge delivery, performance assessment etc. in the \textit{reactive} execution phase. We outline and analyze the major research directions under proactive and reactive engagement of AI in education using a representative group of 194 original research articles published in the past two decades i.e., 2003 - 2022. We discuss the paradigm shifts in the solution approaches proposed, i.e., in the choice of data and algorithms used over this time. We further dive into how the COVID-19 pandemic challenged and reshaped the education landscape at the fag end of this time period. Finally, we pinpoint existing limitations in adopting artificial intelligence for education and reflect on the path forward.
\end{abstract}

% keywords can be removed

\section{Introduction}
In 2015, the United Nations General Assembly identified quality education as one of the seventeen sustainable development goals or SDGs \cite{united2015goal}. The target is to ensure that by 2030, issues pertaining to access to primary, secondary, technical, vocational and tertiary education are addressed globally. In response to this goal, countries have set individual targets in accordance with economic and social development needs. For instance, the United States Department of Education in 2016 adopted a vision for progress in STEM education by 2026 \cite{tanenbaum2016stem}. In a different part of the world, the Ministry of Education in India has rolled out several initiatives to accelerate equitable access to education \cite{ministry2022initiatives}. In this context, it is anticipated globally that technology and more importantly artificial intelligence (AI) driven tools will be central to achieving the holistic goal set by the United Nations General Assembly \cite{vincent2020trustworthy}. 

In the past there has been considerable discourse about how adoption of artificial intelligence driven methods for education might alter the course of how we perceive education \cite{dreyfus1999anonymity,feenberg2017online}. However, in many of the earlier debates, the full potential of artificial intelligence was not recognized due to lack of supporting infrastructure. It was not until very recently that  AI-powered techniques could be used in classroom environments. Since the beginning of the 21st century, there has been a rapid progress in the semiconductor industry in manufacturing chips that can handle computations at scale efficiently. In fact, in the coming decade too it is anticipated that this growth trajectory will continue with focus on wireless communication, data storage and computational resource development \cite{burkacky2022the}. With this ongoing progress, therefore, plans to utilize AI driven tools to support students, educators and policy-makers in education appears to be the logical next step. 

\begin{figure}[h]
  \centering
  \includegraphics[width= \textwidth]{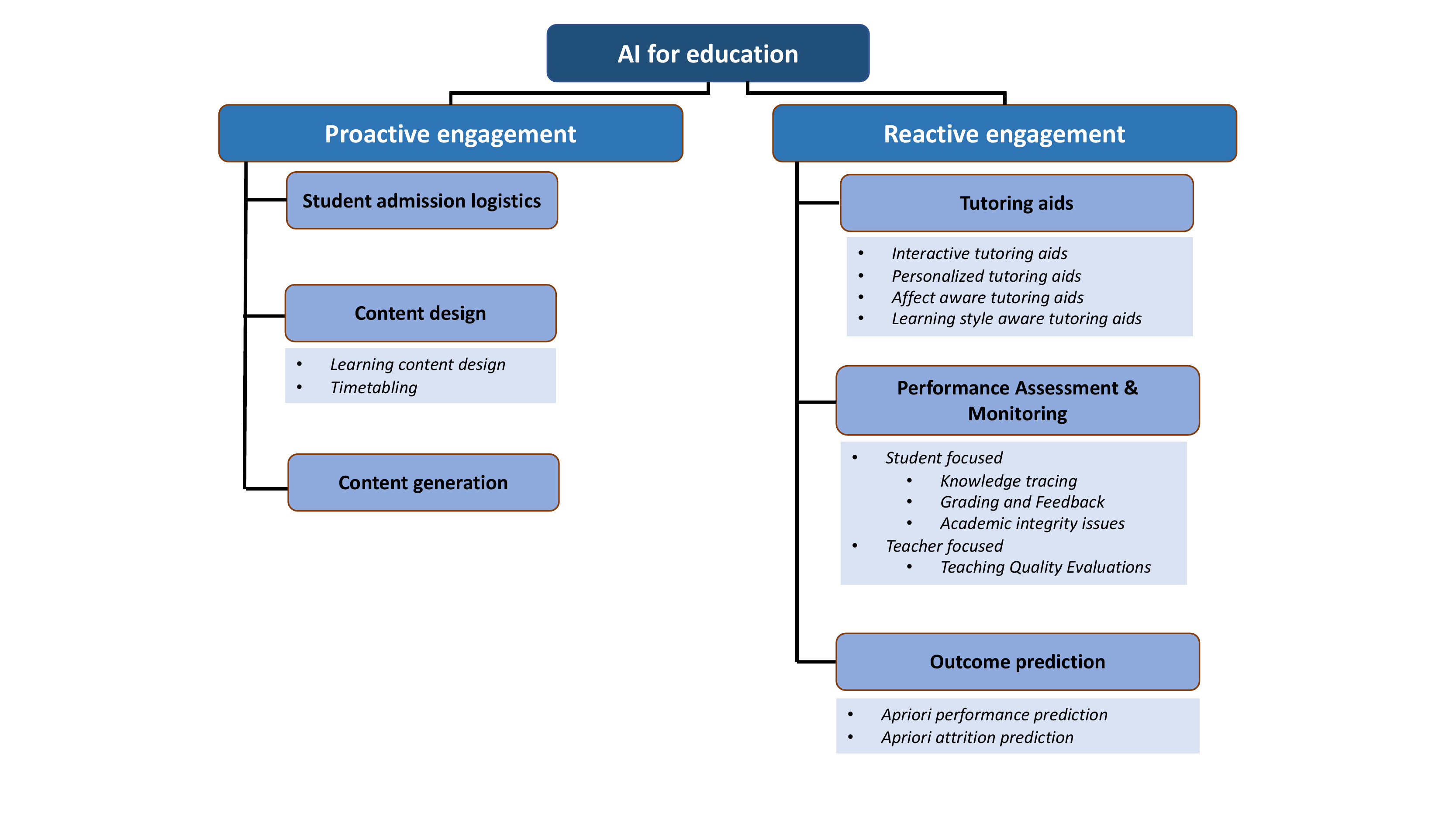}
  \caption{Overview of the categorization introduced in this review article}
  \label{summary_categorization}
\end{figure}

In this review article, we systematically review \textbf{how machine learning and artificial intelligence can be utilized in different phases of the educational process - from planning and scheduling to knowledge delivery and assessment.} To this end, we introduce a broad categorization of original research articles in the literature into methods that are relevant prior to knowledge delivery and those that are relevant in the process of knowledge delivery, i.e., \textit{proactive vs. reactive engagement}. Proactive involvement of AI in education comes from its use in student admission logistics, curriculum design, scheduling, teaching content generation, etc. Reactive involvement of AI is considerably broader in scope - AI-based methods can be used for designing intelligent tutoring systems, assessing performance, predicting student outcomes etc. In the schematic in Fig. \ref{summary_categorization}, we present an overview of our  categorization approach. We have selected a sample set of research articles under each category and identified the key problem statements addressed using AI methods in the past 20 years. 

The COVID-19 pandemic has been one of the most significant social disruptions in recent history. With the outbreak of the virus, many brick-and-mortar educational institutions had to switch into alternate methods of delivering knowledge to students. This in some situations necessitated creative thinking by the administration, educators and students \cite{leblanc2020covid}, and in turn accelerated the adoption and use of technology including artificial intelligence for education. In this article, we highlight how the outbreak of the pandemic globally impacted and shaped the trajectory of artificial intelligence research for education (AIEd). 

Through this review article, we aim to address the following questions:
\begin{itemize}
    \item What were the prominent research directions in involvement of AI in the end-to-end education process in the past two decades? How has the AI methodologies (i.e., choice of datasets and algorithms) evolved over this period in these major research directions?

    \item How did the COVID-19 pandemic influence the education landscape and how AI in particular can drive future developments for educational technologies? 

    \item Does use of AI-driven methodologies for education widen or bridge the gap between population groups when it comes to access to quality education?

\end{itemize}

The organization of this review article from here on is such. In Section 2, we state our search strategy in identifying research articles and present the summary statistics of the articles considered. Here, we clearly define the technical scope of the review. In Section 3, we contextualize our contribution in the light of technical review articles published in the domain of AIEd in the past five years. In Section 4, we present our categorization approach and review the scientific and technical contributions each category. In Section 5, we discuss the impact of the COVID-19 pandemic on AIEd research. Finally, in Section 6, we discuss the existing limitations in the global adoption of AI driven tools for education and the next steps for this field in general. 

\section{Scope Definition}

The term artificial intelligence (AI) was coined in 1956 by John McCarthy \cite{haenlein2019brief}. Since the first generally acknowledged work of McCulloch and Pitts in conceptualizing artificial neurons, AI has gone through several dormant periods and shifts in research focus. More recently, researchers and social scientists are increasingly using AI-based techniques to address social issues and to build towards a sustainable future \cite{shi2020artificial}. In this article, we review the scope of using artificial intelligence to ensure quality education. 

\subsection{Paper Search Strategy}

For the purpose of analyzing recent trends in this field (i.e., AIEd), we have sampled research articles published in peer-reviewed conferences and journals over the past 20 years, i.e. between 2003-2022, by leveraging the Google Scholar search engine using keywords specific to each of our identified use cases. Our aim here was to identify \textit{a sufficient number of research articles that outlined the breadth of use cases using AI-driven methods in planning (i.e., proactive) and execution (i.e., reactive) phases of the education process}. To accomplish this, we did not restrict our search space by considering only certain conferences or journals \cite{shi2020artificial} or by only including articles from authors having a certain h-index \cite{chen2020artificial} or on the basis of citations \cite{chen2020application}. With this strategy in place, we have selected 194 relevant articles. It is worthwhile to mention that this body of 194 articles is by no means the exhaustive list of original research approaches for these use cases, but it is indeed a representative sample size that allows us to identify core research directions in the past two decades and make conclusive statements.

\subsection{Inclusion and Exclusion Criteria}

There is considerable debate in the scientific community about what is the scope of artificial intelligence \cite{fetzer1990artificial}. Here, we do not provide a perspective for what can be included under purview of AI in the context of education, but rather clearly delineate our inclusion/exclusion criteria. For this review article, we include research articles that use methods such as optimal search strategies (eg. BFS, DFS etc.), density estimation, machine learning, Bayesian machine learning, deep learning, reinforcement learning etc. We do not include original research that proposes use of concepts and methods rooted in operations research, evolutionary algorithms, adaptive control theory, robotics etc. in our corpus of selected articles. In this review, we \textbf{only} consider peer-reviewed articles that were published in English. We do not include patented technologies and copyrighted EdTech software systems in our scope unless peer-reviewed articles outlining the same contributions have been published by the authors.  

 \subsection{Summary statistics}

\begin{figure}[h]
  \centering
  \includegraphics[width= 0.8\textwidth]{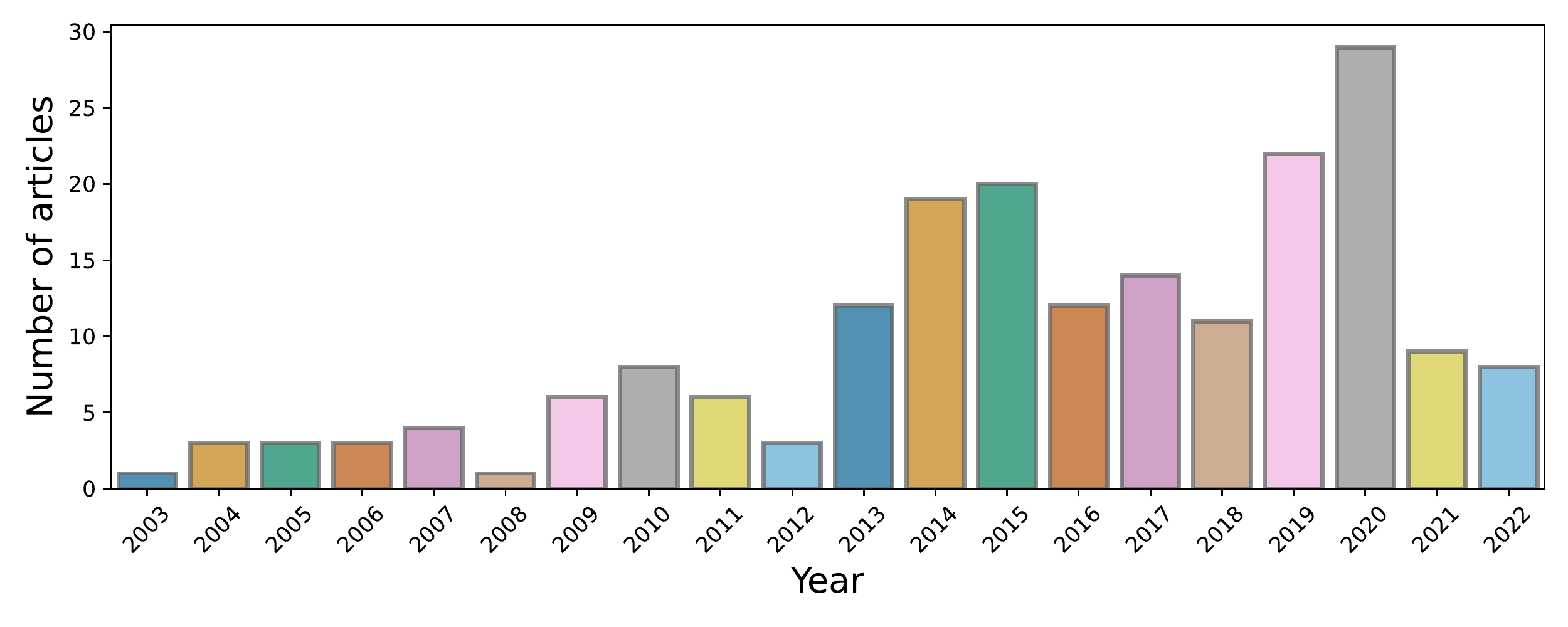}
  \caption{Distribution of the reviewed technical articles across the past two decades}
  \label{count_by_year}
\end{figure}

\begin{figure}[h]
  \centering
  \includegraphics[width= 0.8\textwidth]{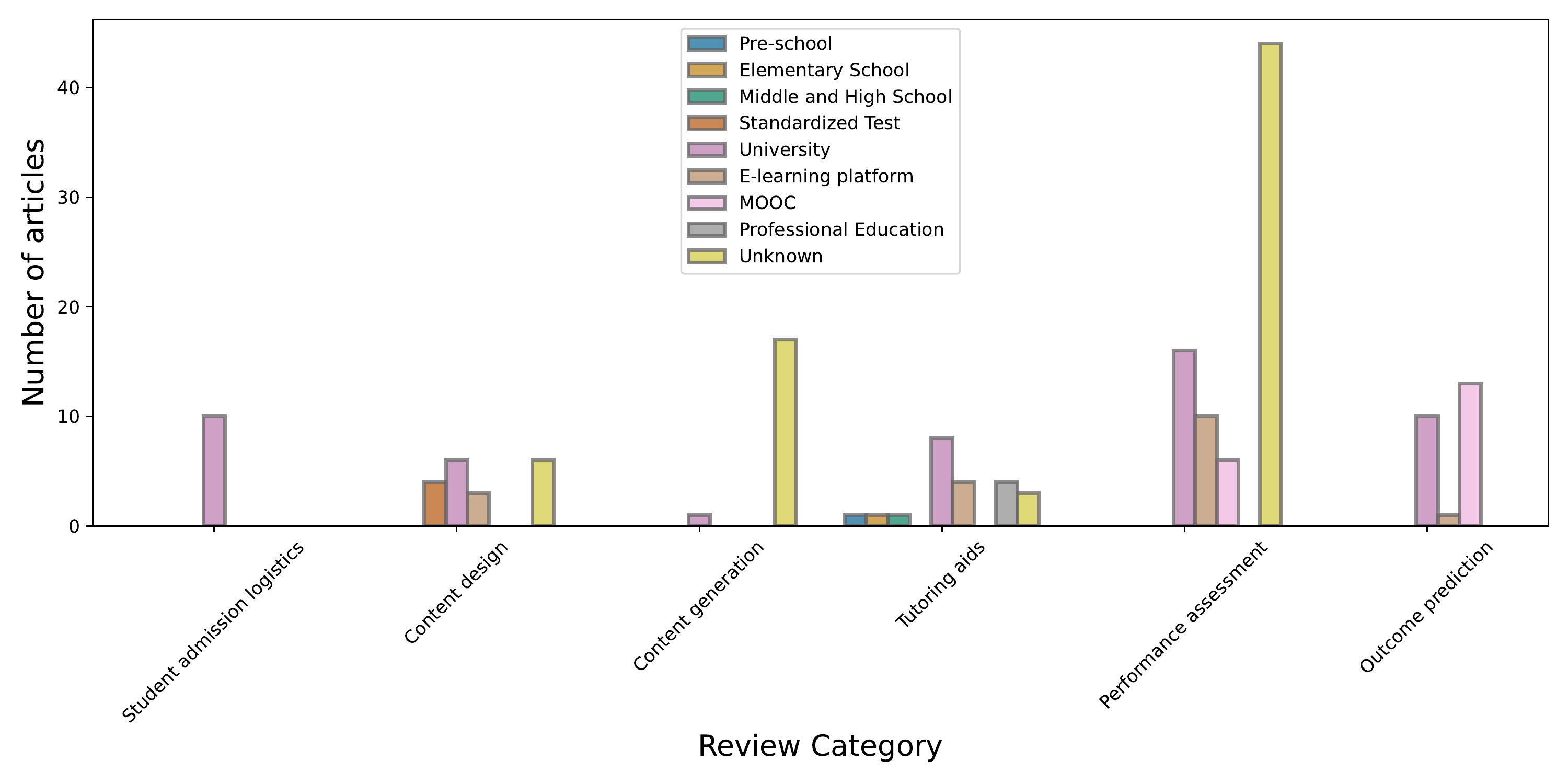}
  \caption{Distribution of reviewed technical articles across categories and target audience categories}
  \label{across_target_review}
\end{figure}

With the scope of our review defined above, here we provide the summary statistics of the 194 technical articles we covered in this review. In Figure \ref{count_by_year}, we show the distribution of the included scientific and technical articles over the past two decades. 
We also introspected the technical contributions in each category of our categorization approach with respect to the target audiences they catered to (see Figure \ref{across_target_review}). We primarily identify target audience groups for educational technologies as such - pre-school students, elementary school students, middle and high school students, university students, standardized test examinees, students in e-learning platforms, students of MOOCs, and students in professional/vocational education. Articles where the audience group has not been clearly mentioned were marked as belonging to `Unknown` target audience category.

\begin{figure}[h]
  \centering
  \includegraphics[width= \textwidth]{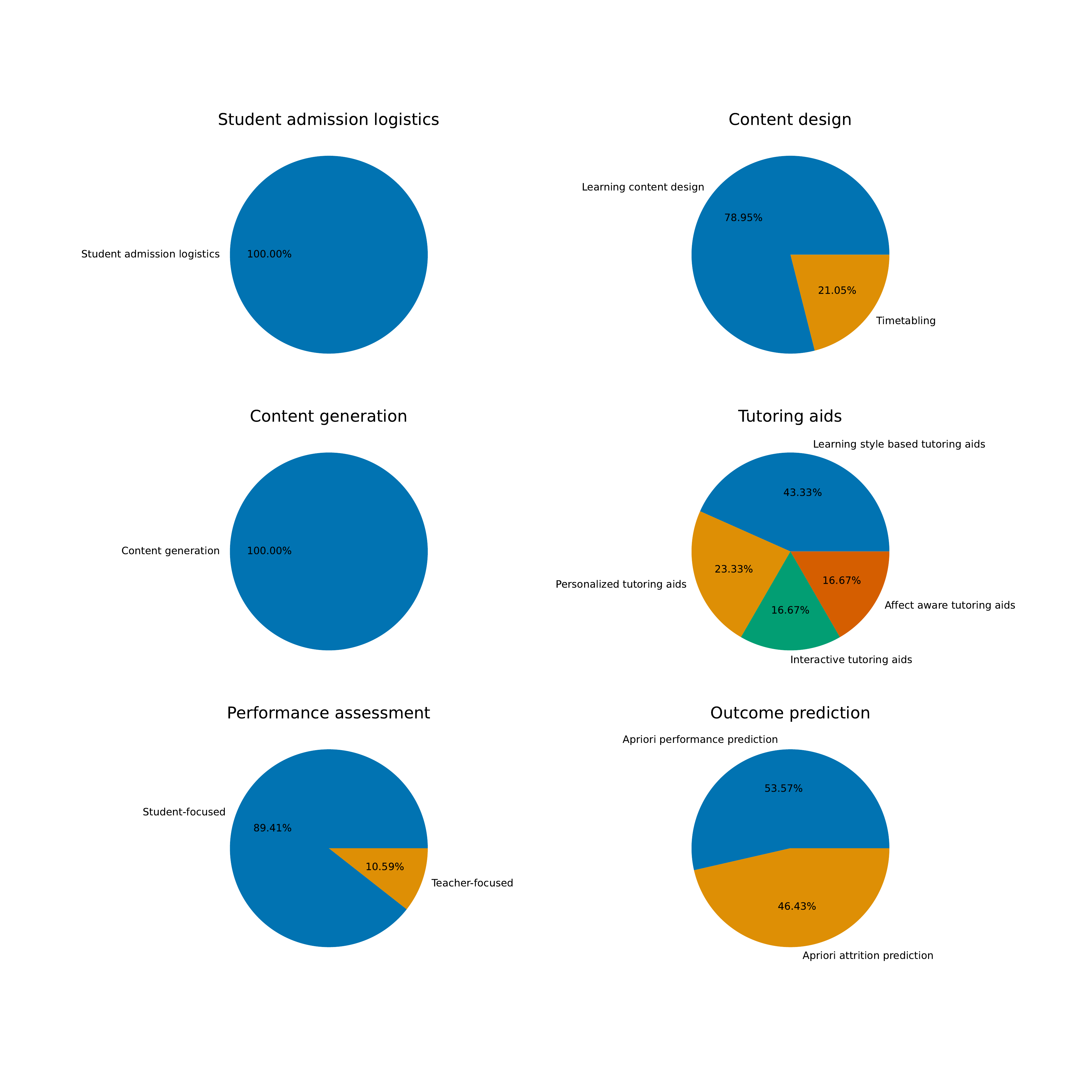}
  \caption{Distribution of reviewed technical articles across sub-categories under each category}
  \label{subcategories}
\end{figure}

In Section 4, we introduce our categorization and perform a deep-dive to explore the breadth of technical contributions in each category. If applicable, we have further identified specific research problems currently receiving much attention as sub-categories within a category. In Figure \ref{subcategories}, we demonstrate the distribution of significant research problems within a category. 

We defer the analysis of the identified trends from these summary plots to the Discussion section of this paper. 

\section{Related Works}

\begin{figure}[h]
  \centering
  \includegraphics[width= 0.8\textwidth]{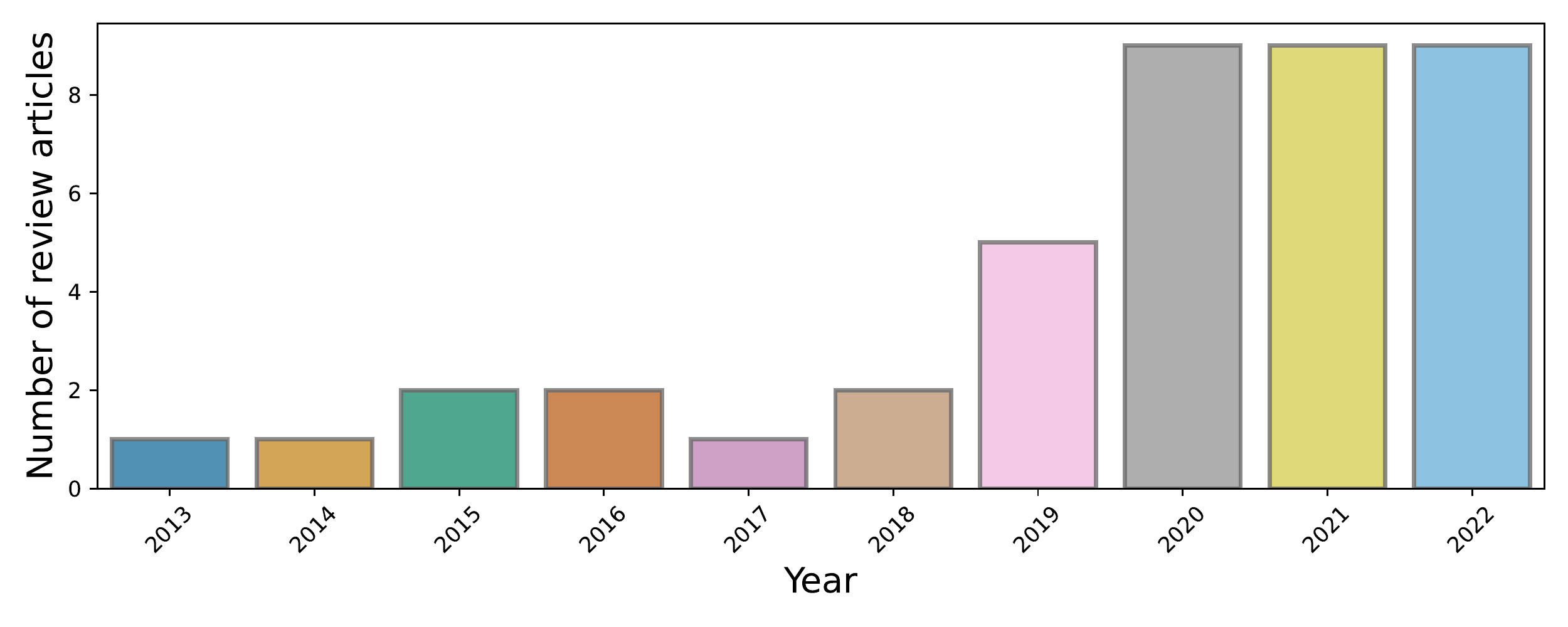}
  \caption{Number of review articles published in AIEd over the past decade.}
  \label{review_articles_vs_year}
\end{figure}

Since being identified as one among the seventeen SDGs \cite{united2015goal}, there has been increased attention in addressing existing issues in the field of education through the latest and greatest technology. It must be noted however, neither the use of technology to benefit education nor artificial intelligence as a technology is an invention of the twenty-first century. It is in fact the increased social awareness about equitable access to developmental resources and simultaneous development of the infrastructure to support it that led to the increase in scientific and technical contributions in AIEd research. In this backdrop, the number of review articles surveying the technical progress in this discipline has also increased in the last decade (see Fig. \ref{review_articles_vs_year} - \textit{note that we used Google Scholar as the search engine with the keywords artificial intelligence for education, artificial intelligence for education review articles etc.}). Here, we discuss the premise of the review articles published in the \textit{last five years} and contextualize this article with respect to previously published technical reviews. 

Among the review articles identified based on the keyword search on Google Scholar and published between 2018 and 2022, one can identify two thematic categories - (i) \textit{Technical reviews with categorization}: review articles that group research contributions based on some distinguishing factors such as problem statement, solution methodology etc \cite{chassignol2018artificial, zawacki2019systematic, chen2020artificial, yufeia2020review, ahmad2020artificial, zhai2021review, ouyang2021artificial, lameras2021power, huang2021review}. (ii) \textit{Perspectives on challenges, trends and roadmap}: review articles that highlight the current state of research in a domain and offer critical analysis of the challenges and the future road map for the domain \cite{malik2019analysis, fahimirad2018review, humble2019artificial, pedro2019artificial, hwang2020vision, schiff2021out}. Closely linked with (i) are review articles that dive deep into the developments within a particular sub-category associated with AIEd, such as AIEd in the context of early childhood education \cite{su2022artificial}, online higher education \cite{ouyang2022artificial} etc. We have designed this review article to belong to category (i) - we distinguish between the use of artificial intelligence for education based on their proactive or reactive involvement in the education process. To the best of our knowledge, we for the first time categorize AIEd research articles through such lens and provide an in-depth review of significant research problems in each category (see schematic in Fig. \ref{summary_categorization}). We believe that our categorization approach introduces researchers to the wide scope of using AI-driven methods for providing quality education. At the same time, the article summarizes for expert researchers the progress of AI research in utilizing diverse datasets and algorithms for problem statements in the education sector and the scope for future research. 

In Table \ref{contextualization}, we have outlined the context of recent categorical review articles along with ours to provide readers a comprehensive summary of how we place this article in the body of literature for AIEd. 

\begin{longtable}{|p{5cm}|p{2cm}|p{1cm}|p{7cm}|}

\caption{\textbf{Contextualization with respect to technical reviews published in the past five years (2018 - 2022)}}
\label{contextualization} \\

\hline
\hline
\textbf{Paper title} & \textbf{Authors} & \textbf{Year} & \textbf{Summary} \\
\hline
Artificial Intelligence trends in education: a narrative overview \cite{chassignol2018artificial} & Chassignol et. al. & 2018  & Categorizes AI in education into four categories - customized educational content, assessment and evaluation, adaptive systems and personalization, intelligent tutoring systems. \\

\hline
Systematic review of research on artificial intelligence applications in higher education - where are the educators \cite{zawacki2019systematic} & Zawacki-Richter et. al. & 2019  & Categorizes AI in education into four categories - profiling and prediction, assessment and evaluation, adaptive systems and personalization, intelligent tutoring systems. \\

\hline
Artificial Intelligence in Education: A Review \cite{chen2020artificial} & Chen et. al. & 2020  & Identifies and reviews four key ways in which AI has been adopted for education - automation of administrative processes and tasks, curriculum and content development, instruction, modeling students' learning process. \\

\hline
Review of the application of artificial intelligence in education \cite{yufeia2020review} & Yufei et. al.  & 2020  & Identifies and reviews aspects in which AI technology has been used in education - automatic grading system, interval reminder, teacher’s feedback, virtual teachers, personalized learning, adaptive learning, augmented reality/virtual reality, accurate reading, intelligent campus and distance learning.  \\

\hline
Artificial Intelligence in Education: A panoramic review \cite{ahmad2020artificial} & Ahmed et. al. & 2020  & Reviews the various applications of AI such as student grading and evaluations, students retention and drop out prediction, sentiment analysis, intelligent tutoring, classroom monitoring and recommendation systems. \\

\hline
A Review of Artificial Intelligence (AI) in Education from 2010 to 2020 \cite{zhai2021review} & Zhai et.al. & 2021 & Reviews articles that use AI for social sciences such as in education and classifies the research questions into development layer (classification, matching, recommendation and deep learning), application layer (feedback, reasoning and adaptive learning), and integration layer (affection computing, role-playing, immersive learning and gamification).\\

\hline
Artificial intelligence in education: The three paradigms\cite{ouyang2021artificial} & Ouyang et. al. & 2021 & Identifies the paradigm shifts of AIEd and categorizes into AI-directed (learner-as-recipient), AI-supported (learner-as-collaborator), and AI-empowered (learner-as-leader). \\

\hline
Power to the teachers: an exploratory review on artificial intelligence in education\cite{lameras2021power} & Lameras et. al. & 2021 & Discusses research contribution along the five aspects of teaching and learning introduced by \cite{dong2020seldon} - 1. AIEd for preparing and transmitting learning content 2. AIEd for helping students to apply knowledge 3. AIEd for engaging students in learning tasks 4. AIEd for helping students to improvement through assessments and feedback 5. AIEd for helping students to become self-regulated learners. \\

\hline
A review on artificial intelligence in education\cite{huang2021review} & Huang et. al. & 2021 & Outlines the application of AI in education - adaptive learning, teaching evaluation, virtual classroom, smart campus, intelligent tutoring robots and then analyzes its impact on teaching and learning. \\

\hline
Towards a tripartite research agenda: A scoping review of artificial intelligence in education research \cite{wang2022towards} &  Wang et. al. & 2022 & Provides a scoping review of research studies on AIEd published between 2001–2021 and identifies and discusses three distinct agendas - Learning from AI, Learning about AI, and Learning with AI. \\

\hline
Two Decades of Artificial Intelligence in Education: Contributors, Collaborations, Research Topics, Challenges, and Future Directions \cite{chen2022two} & Chen et. al. & 2022 & The authors identify the main research topics in AIEd in the past two decades to be - intelligent tutoring systems for special education, natural language processing for language education, educational robots for AI education,  educational data mining for performance prediction, discourse analysis in computer-supported collaborative learning, neural networks for teaching evaluation, affective computing for learner emotion detection, and recommender systems for personalized learning. \\

\hline
Academic and Administrative Role of Artificial Intelligence in Education \cite{ahmad2022academic} & Ahmad et. al. & 2022 & This review article aims to explore the academic and administrative applications of AI with an in-depth discussion on artificial intelligence applications in 1. Grading/Assessment 2. Admission 3. Virtual Reality (VR) for education 4. Learning Analytics etc. \\

\hline
A Comprehensive Overview of Artificial Intelligence Trends in Education \cite{namatherdhala2022comprehensive} & Namatherdhala et. al. & 2022 & The authors categorize application of AI for education into three distinct groups - Education administration, Instruction Design and Learning outcomes and briefly reviews each of them. \\

\hline
\textbf{Proactive and Reactive engagement of Artificial Intelligence methods for Education: A review } & \textbf{Mallik \& Gangopadhyay} &  & \textbf{We introduce a novel categorization that distinguishes between proactive (admissions, scheduling, content generation etc.) and reactive (tutoring systems, performance assessment, outcome prediction etc.) involvement of AI for education and review scientific and technical contributions under each sub-category over the past two decades.}\\

\hline
\hline

\end{longtable}

\section{Engaging artificial intelligence driven methods in stages of education }

\subsection{Proactive vs Reactive engagement of AI - an introduction}

The process of educating a student begins much before the student starts attending lectures and parsing lecture materials. In a traditional education setup, administrative staff and educators begin preparations related to making admissions decisions, scheduling of classes to optimize resources, curating course contents and preliminary assignment materials several weeks prior to the term start date. Once the term starts, the focus of educators is to deliver the course material, give out and grade assignments to assess progress, and provide additional support to students who might benefit from that. The role of the students is to regularly acquire knowledge, ask clarifying questions and seek help to master the material. The role of administrative staff in this phase is less hands-on - they remain involved to ensure smooth and efficient overall progress. Therefore, we can clearly identify two distinct work phases in the end-to-end education process. First, \textbf{proactive engagement} - all efforts in this phase are to design and curate to ensure optimal use of resources, and second, \textbf{reactive engagement} - all efforts in this phase are to ensure that students acquire the necessary information and skills from the sessions they attend and to address any blockers they might encounter. In this review article, we leverage these two phases to distinguish between contributions in the field of AIEd research. Our primary categorization of AIEd research is therefore, (i) \textbf{Proactive engagement of AI for education}, and (ii) \textbf{Reactive engagement of AI for education}. Within these broad categorizations, we further identify different genres of research relevant to the education sector. For instance, in the proactive engagement phase, AI-based algorithms can be leveraged to determine student admission logistics, design curricula and schedules, and create course content. On the other hand, in the reactive engagement phase, AI-based methods can be used for designing intelligent tutoring systems (ITS), performance assessment, prediction of student outcomes etc. (see Figure 1). Another important distinction between the two phases lies in the nature of the available data to develop models. While the former primarily makes use of historical data points or pre-existing estimates of available resources and expectations about learning outcomes, the latter has at its disposal a growing pool of data points from the currently ongoing learning process, and can therefore be more adaptive and initiate faster pedagogical interventions to changing scopes and requirements.

\subsection{Proactive engagement of AI for education}

\subsubsection{Student admission logistics} 

In the past, although a number of studies used statistical or machine learning-based approaches to analyze or model student admissions decisions, they had little role in the actual admissions process \cite{bruggink1996statistical,moore1998expert}. However in the face of growing numbers of applicants, educational institutes are increasingly turning to AI-driven approaches to efficiently review applications and make admission decisions. For example, the Department of Computer Science at University of Texas Austin (UTCS) introduced an explainable AI system called GRADE (Graduate Admissions Evaluator) that uses logistic regression on past admission records to estimate the probability of a new applicant being admitted in their graduate program \cite{waters2014grade}. While GRADE did not make the final admission decision, it reduced the number of full application reviews as well as review time per application by experts. \cite{zhao2020quantitative} used features extracted from application materials of students as well as how they performed in the program of study to predict an incoming applicant's potential performance and identify students best suited for the program.
An important metric for educational institutes with regard to student admissions is yield rate, the rate at which accepted students decide to enroll at a given school. Machine learning has been used to predict enrollment decisions of students, which would help the institute make strategic admission decisions in order to improve their yield rate and optimize resource allocation \cite{jamison2017applying,esquivel2020using}. Additionally, whether students enroll in suitable majors based on their specific backgrounds and prior academic performance is also indicative of future success. Machine learning has also been used to classify students into suitable majors in an attempt to set them up for academic success \cite{assiri2022improve}.

Another research direction in this domain approaches the admissions problem from the perspective of students by predicting the probability that an applicant will get admission at a particular university in order to help applicants better target universities based on their profiles as well as university rankings \cite{alghamdi2020machine,goni2020graduate,mridha2022machine}. Notably, more than one such work finds prior GPA (Grade Point Average) of students to be the most significant factor in admissions decisions \cite{young2019using,el2021recommender}.

Given the high stakes involved and the significant consequences that admissions decisions have on the future of students, there has been considerable discourse on the ethical considerations of using AI in such applications, including its fairness, transparency and privacy aspects \cite{agarwal2020trade,finocchiaro2021bridging}. Aside from the obvious potential risks of worthy applicants getting rejected or unworthy applicants getting in, such systems can perpetuate existing biases in the training data from human decision-making in the past \cite{bogina2022educating}. For example, such systems might show unintentional bias towards certain demographics, gender, race, income groups, etc. \cite{bogina2022educating} advocated for explainable models for making admission decisions, as well as proper system testing and balancing before reaching the end user. \cite{emelianov2020fair} showed that demographic parity mechanisms like group-specific admission thresholds increase the utility of the selection process in such systems in addition to improving its fairness. Despite concerns regarding fairness and ethics, interestingly, university students in a recent survey rated algorithmic decision-making (ADM) higher than human decision-making (HDM) in admission decisions in both procedural and distributive fairness aspects \cite{marcinkowski2020implications}.

\subsubsection{Content design}
In the context of education, we can define content 
as - (i) learning content for a course, curriculum or test; and (ii) schedules/timetables of classes. We discuss AI/ML approaches for both of the above in this section.

(i) \textbf{Learning content design}: Prior to the start of the learning process, educators and administrators are responsible for identifying an appropriate set of courses for a curriculum, 
an appropriate set of contents for a course, or an appropriate set of questions for a standardized test. In course and curriculum design, there is a large body of work using traditional systematic and relational approaches \cite{kessels1999relational}, however the last decade saw several works using AI-informed curriculum design approaches. For example, \cite{ball2019applying} uses classical ML algorithms to identify factors prior to declaration of majors in universities that adversely affect graduation rates, and advocates curriculum changes to alleviate these factors. 
\cite{rawatlal2017application} uses tree-based approaches on historical records to prioritize the prerequisite structure of a curriculum in order to determine student progression routes that are effective.
\cite{somasundaram2020curriculum} proposes an Outcome Based Education (OBE) where expected outcomes from a degree program such as job roles/skills are identified first, and subsequently courses required to reach these outcomes are proposed by modeling the curriculum using ANNs. \cite{doroudi2019integrating} suggests a semi-automated curriculum design approach by automatically curating low-cost, learner-generated content for future learners, but argues that more work is needed to explore data-driven approaches in curating pedagogically useful peer content.

For designing standardized tests such as TOEFL, SAT or GRE, an essential criteria is to select questions having a consistent difficulty level across test papers for fair evaluation. This is also useful in classroom settings if teachers want to avoid plagiarism issues by setting multiple sets of test papers, or in designing a sequence of assignments or exams with increasing order of difficulty. This can be done through Question Difficulty Prediction (QDP) or Question Difficulty Estimation (QDE), an estimate of the skill level needed to answer a question correctly. QDP was historically estimated by pretesting on students or from expert ratings, which are expensive, time-consuming, subjective and often vulnerable to leakage or exposure \cite{benedetto2022survey}.
Rule-based algorithms relying on difficulty features extracted by experts were also proposed in \cite{grivokostopoulou2014teaching,perikos2016automatic} for automatic difficulty estimation.
As data-driven solutions became more popular, a common approach used linguistic features \cite{mothe2005linguistic,stiller2016assessing}, readability scores, \cite{benedetto2020introducing,yaneva2020predicting} and/or word frequency features 
\cite{benedetto2020introducing,yaneva2020predicting,benedetto2020r2de} with ML algorithms such as linear regression, SVMs, tree-based approaches, neural networks, etc. for downstream classification or regression, depending on the problem setup.
With automatic testing systems and ready availability of large quantities of historical test logs, deep learning has been increasingly used for feature extraction (word embeddings, question representations, etc.) and/or difficulty estimation \cite{fang2019exercise,lin2019automated,xue2020predicting}. Attention strategies have been used to model the difficulty contribution of each sentence in reading problems \cite{huang2017question} or to model recall (how hard it is to recall the knowledge assessed by the question) and confusion (how hard it is to separate the correct answer from distractors) in \cite{qiu2019question}. Domain adaptation techniques have also been proposed to alleviate the need of difficulty-labeled question data for each new course by aligning it with the difficulty distribution of a resource-rich course \cite{huang2021stan}. \cite{alkhuzaey2021systematic} points out that a majority of data-driven QDP approaches belong to language learning and medicine, possibly spurred on by the existence of a large number of international and national-level standardized language proficiency tests and medical licensing exams.

(ii) \textbf{Timetabling}: Educational Timetabling Problem (ETP) deals with the assignment of classes or exams to a limited number of time-slots such that certain constraints (e.g. availability of teachers, students, classrooms, equipments, etc.) are satisfied. This can be divided into three types - course timetabling, school timetabling and exam timetabling \cite{zhu2021survey}. 
Timetabling not only ensures proper resource allocation, its design considerations (e.g. number of courses per semester, number of lectures per day, number of free time-slots per day, etc.) have noticeable impact on student attendance behavior and academic performance \cite{larabi2021impact}. 
%Moreover with a growing number of students and accompanying constraints in learning settings, automating timetabling systems to make best use of finite resources have become critical. 
Popular approaches in this domain such as mathematical optimization, meta-heuristic, hyper-heuristic, hybrid, fuzzy logic approaches, etc. \cite{zhu2021survey,tan2021survey} mostly is beyond the scope of our paper (see Section 2.2). Having said that, it must be noted that machine learning has often been used in conjunction with such mathematical techniques to obtain better performing algorithms. 
For example, \cite{kenekayoro2020incorporating} used supervised learning to find approximations for evaluating solutions to optimization problems - a critical step in heuristic approaches. Reinforcement learning has been used to select low-level heuristics in hyper-heuristic approaches \cite{obit2011non,ozcan2012reinforcement} or to obtain a suitable search neighborhood in mathematical optimization problems \cite{goh2019simulated}.

\subsubsection{Content generation}

The difference between content design and content generation is that of curation versus creation. While the former focuses on selecting and structuring the contents for a course/curriculum in a way most appropriate for achieving the desired learning outcomes, the latter deals with generating the course material itself. AI has been widely adopted to generate and improve learning content prior to the start of the learning process, as discussed in this section.

Automatically generating questions from narrative or informational text, or automatically generating problems for analytical concepts are becoming increasingly important in the context of education. Automatic question generation (AQG) from teaching material can be used to improve learning and comprehension of students, assess information retention from the material and aid teachers in adding supplemental material from external sources without the time-intensive process of authoring assessments from them. They can also be used as a component in intelligent tutoring systems to drive engagement and assess learning.
%As pointed out in \cite{das2021automatic}, a majority of work in this domain focuses on generating objective-type or close-ended questions such as true/false, matching, fill-in-the-blank (cloze) or multiple choice questions (MCQ), in which a critical component involves generating distractors (incorrect answer choices). 
AQG essentially consists of two aspects: content selection or \textit{what to ask}, and question construction or \textit{how to ask it} \cite{pan2019recent}, traditionally considered as separate problems. Content selection for questions was typically done using different statistical features (sentence length, word/sentence position, word frequency, noun/pronoun count, presence of superlatives, etc.) \cite{agarwal2011automatic} or NLP techniques such as syntactic or semantic parsing \cite{heilman2011automatic,lindberg2013generating}, named entity recognition \cite{kalady2010natural}, topic modeling \cite{majumder2015system} etc. 
Machine learning has also been used in such contexts, e.g. to classify whether a certain sentence is suitable to be used as a stem in cloze questions (passage with a portion occluded which needs to be replaced by the participant) \cite{correia2012automatic}. The actual question construction, on the other hand, traditionally adopted rule-based methods like transformation-based approaches \cite{varga2010wlv} or template-based approaches \cite{mostow2009generating}. The former rephrased the selected content using the correct question key-word after deleting the target concept, while the latter used pre-defined templates that can each capture a class of questions. \cite{heilman2010good} used an overgenerate-and-rank approach to overgenerate questions followed by the use of supervised learning for ranking them, but still relied on handcrafted generating rules. Following the success of neural language models and concurrent with the release of large-scale machine reading comprehension datasets \cite{rajpurkar2016squad,nguyen2016ms}, question generation was later framed as a sequence-to-sequence learning problem that directly maps a sentence (or the entire passage containing the sentence) to a question \cite{du2017learning,zhao2018paragraph,kim2019improving}, and can thus be trained in an end-to-end manner \cite{pan2019recent}. 
Reinforcement learning based approaches that exploit the rich structural information in the text have also been explored in this context \cite{chen2019reinforcement}.
While text is the most common type of input in AQG, such systems have also been developed for structured databases \cite{jouault2013building,indurthi2017generating}, images \cite{mostafazadeh2016generating} and videos \cite{huang2014tedquiz}, and are typically evaluated by experts on the quality of generated questions in terms of relevance, grammatical and semantic correctness, usefulness, clarity etc.

Automatically generating problems that are similar to a given problem in terms of difficulty level, can greatly benefit teachers in setting individualized practice problems to avoid plagiarism and still ensure fair evaluation \cite{ahmed2013automatically}. It also enables the students to be exposed to as many (and diverse) training exercises as needed in order to master the underlying concepts \cite{keller2021automatic}. In this context, mathematical word problems (MWPs) - an established way of inculcating math modeling skills in K-12 education - have witnessed significant research interest. Preliminary work in automatic MWP generation take a template-based approach, where an existing problem is generalized into a template, and a solution space fitting this template is explored to generate new problems \cite{deane2003automatic,polozov2015personalized,koncel2016theme}. Following the same shift as in AQG, \cite{zhou2019towards} proposed an RNN-based approach that encodes math expressions and topic words to automatically generate such problems. Subsequent research along this direction has focused on improving topic relevance, expression relevance, language coherence, as well as completeness and validity of the generated problems using a spectrum of approaches \cite{liu2020mathematical,wang2021math,wu2022automatic}.

On the other end of the content generation spectrum lie systems that can generate solutions based on the content and related questions, which include Automatic Question Answering (AQA) systems, Machine Reading Comprehension (MRC) systems and automatic quantitative reasoning problem solvers \cite{zhang2019gap}. These have achieved impressive breakthroughs with the research into large language models and are widely regarded in the larger narrative as a stepping-stone towards Artificial General Intelligence (AGI), since they require sophisticated natural language understanding and logical inferencing capabilities. However their applicability and usefulness in educational settings remains to be seen.

\subsection{Reactive engagement of AI for education}

\subsubsection{Tutoring aids}

Technology has been used to aid learners to achieve their learning goals for a long time. More focused effort on developing computer-based tutoring systems in particular started following the findings of Bloom \cite{bloom19842} - students who received tutoring in addition to group classes fared two standard deviations better than those who only participated in group classes. Given its early start, research on Intelligent Tutoring Systems (ITS) is relatively more mature than other research areas under the umbrella of AIEd research. Fundamentally, the difference between designs of ITS comes from the difference in the \textit{underlying assumption of what augments the knowledge acquisition process for a student.} In the review paper on ITS \cite{alkhatlan2018intelligent}, a comprehensive timeline and overview of research in this domain is provided. Instead of repeating findings from previous reviews under this category, we distinguish between ITS designs through the lens of the underlying hypotheses. We primarily identified four hypotheses that are currently receiving much attention from the research community - emphasis on tutor-tutee interaction, emphasis of personalization, inclusion of affect and emotion, and consideration of specific learning styles. It must be noted that tutoring itself is an interactive process, therefore most designs in this category have a basic interactive setup. However, contributions in categories (ii) through (iv), have other concept as the focal point of their tutoring aid design. 

(i) \textbf{Interactive tutoring aids}: Previous research in education \cite{jackson2013motivation} has pointed out that \textit{when a student is actively interacting with the educator or the course contents, the student stays engaged in the learning process for a longer duration}. Learning systems that leverage this hypothesis can be categorized as interactive tutoring aids. These frameworks allow the student to communicate (verbally or through actions) with the teacher or the teaching entity (robots or software) and get feedback or instructions as needed.  

Early designs of interactive tutoring aids for teaching and support comprised of rule-based systems mirroring interactions between expert teacher and student \cite{arroyo2004wayang,olney2012guru} or between peer companions \cite{movellan2009sociable}. These template rules provided output based on the inputs from the student. Over the course of time, interactive tutoring systems gradually shifted to inferring the student's state in real time from the student's interactions with the tutoring system and providing fine-tuned feedback/instructions based on the inference. For instance, \cite{gordon2015bayesian} used a Bayesian active learning algorithm to assess student's word reading skills while the student was being taught by a robot. Presently, a significant number of frameworks belonging to this category uses chatbots as a proxy for a teacher or a teaching assistant \cite{ashfaque2020review}. These recent designs can use a wide variety of data such as text, speech etc. and rely on a combination of sophisticated and resource-intensive deep-learning algorithms to infer and further customize interactions with the student. For example, \cite{pereira2016leveraging} presents `@dawebot' that uses NLP techniques to train students using multiple choice question quizzes. \cite{afzal2020ai} presents a conversational medical school tutor that uses NLP and natural language understanding (NLU) to understand user's intent and present concepts associated with a clinical case.  

Hint construction and partial solution generation is yet another method to keep students engaged interactively. For instance,  \cite{green2011learning} used Dynamic Bayes Nets to construct a curriculum of hints and associated problems. \cite{wang2015automated} in their architecture iGeoTutor assisted students in mastering geometry theorems by implementing search strategies (e.g. DFS) from partially complete proofs. \cite{pande2021hybrid} aims to improve individual and self-regulated learning in group assignments through a conversational system built using NLU and dialogue management systems that prompts the students to reflect on lessons learnt while directing them to partial solutions, etc. 

One of the requirements of certain professional and vocational training such as biology, medicine, military etc. is practical experience. With the support of booming infrastructure, many such training programs are now adopting AI-driven augmented reality (AR)/ virtual reality (VR) lesson plans. Interconnected modules driven by computer vision, NLU, NLP, text-to-speech (TTS), information retrieval algorithms facilitate lessons and/or assessments in biology \cite{ahn2018intelligent}, surgery and medicine \cite{mirchi2020virtual}, pathological laboratory analysis \cite{taoum2016design}, military leadership training \cite{gordon2004branching} etc. 

(ii) \textbf{Personalized tutoring aids}: As every student is unique, \textit{personalizing instruction and teaching content can positively impact the learning outcome of the student} \cite{walkington2013using} - tutoring systems that incorporate this can be categorized as personalized learning systems or personalized tutoring aids. Notably, personalization during instruction can occur through course content sequencing, display of prompts and additional resources, etc. 

The sequence in which a student reviews course topics plays an important role in their mastery of a concept. One of the criticisms of early computer based learning tools was the `one approach fits all' method of execution. To improve upon this limitation, personalized instructional sequencing approaches were adopted. In some early developments, \cite{idris2009adaptive} developed a course sequencing method that mirrored the role of an instructor using soft computing techniques such as self organized maps and feed-forward neural networks. \cite{lin2013data} propose the use of decision trees trained on student background information to propose personalized learning paths for creativity learning. Reinforcement learning naturally lends itself to this task. Here an optimal policy (sequence of instructional activities) is inferred depending on the cognitive state of a student (estimated through knowledge tracing) in order to maximize a learning-related reward function. As knowledge delivery platforms are increasingly becoming virtual and thereby generating more data, deep reinforcement learning has been widely applied to the problem of instructional sequencing \cite{reddy2017accelerating, upadhyay2018deep, pu2020deep, islam2021pakes}. \cite{doroudi2019integrating} presents a systematic review of RL-induced instructional policies that were evaluated on students, and concludes that over half outperform all baselines they were tested against.

In order to display a set of relevant resources personalized with respect to a student state, algorithmic search is carried out in a knowledge repository. For instance, \cite{kim2009pedagogical} uses information retrieval and NLP techniques to present two frameworks: PedaBot that allows students to connect past discussions to the current discussion thread and MentorMatch that facilitates student collaboration customized based on student's current needs etc. Both PedaBot and MentorMatch systems use text data coming from a live discussion board in addition to textbook glossaries.  In order to reduce information overload and allow learners to easily navigate e-learning platforms, Deep Learning-Based Course Recommender System (DECOR) has been proposed recently \cite{li2021deep} - this architecture comprises of neural network based recommendation systems trained using student behavior and course related data.

(iii) \textbf{Affect aware tutoring aids}: Scientific research proposes \textit{incorporating affect and behavioral state of the learner into the design of the tutoring system as it enhances the effectiveness of the teaching process} \cite{san2013towards, woolf2009affect}. In \cite{arroyo2014multimedia}, Arroyo et. al. suggests that cognition, meta-cognition and affect should indeed be modeled using real time data and used to design intervention strategies. Affect and behavioral state of a student can generally be inferred from sensor data that tracks minute physical movements of the student (eyegaze, facial expression, posture etc.). While initial approaches in this direction required sensor data, a major constraint for availing and using such data pertains to ethical and legal reasons. `Sensor-free' approaches have thereby been proposed that use data such as student self-evaluations and/or interaction logs of the student with the tutoring system. \cite{woolf2010effect, arroyo2010bayesian} use interaction data to build affect detector models - the raw data in these cases are first distilled into meaningful features and then fed into simple classifier models that detect individual affective states. \cite{defalco2018detecting} compares the usage of sensor and interaction data in delivering motivational prompts in the course of military training. In \cite{botelho2017improving}, uses RNNs to enhance the performance of sensor-free affect detection models. In their review of affect and emotion aware tutoring aids, \cite{harley2017developing} explore in depth the different use cases for affect aware intelligent tutoring aids such as enriching user experience, better curating learning material and assessments, delivering prompts for appraisal, navigational instructions etc and the progress of research in each direction. 

(iv) \textbf{Learning style aware tutoring aids}: Yet another perspective in the domain of ITS is that \textit{understanding the learning styles of the students prior to delivery of course content leads to better end outcomes}. \cite{pask1976styles, kolb1976learning, honey1986manual, felder2002learning} among others proposed different approaches to categorize learning styles of students. Traditionally, an individual’s learning style was inferred via use of a self-administered questionnaire. However, more recently machine learning based methods are being used to categorize learning styles more efficiently from noisy subject data. \cite{lo2005identification, villaverde2006learning, alfaro2018online, bajaj2018smart} use as input the completed questionnaire and/or other data sources such as interaction data, behavioral data of students etc and feed the extracted features into feed-forward neural networks for classification. Unsupervised methods such as self-organizing map (SOM) trained using curated features have also been used for automatic learning style identification \cite{zatarain2010learning}. While for categorization per the Felder and Silverman learning style model, count of student visits to different sections of the e-learning platform are found to be more informative \cite{bajaj2018smart}, \cite{bernard2015using}, for categorization per the Kolb learning model, student performance and student preference features were found to be more relevant.  Additionally, machine learning approaches have also been proposed for learning style based learning path design. In \cite{mota2008using}, learning styles are first identified through a questionnaire and represented on a polar map, thereafter neural networks are used to predict the best presentation layout of the learning objective for a student.  

\subsubsection{Performance assessment and monitoring}

A critical component of the knowledge delivery phase involves assessing student performance by tracing their knowledge development and providing grades and/or constructive feedback on assignments and exams, while simultaneously ensuring academic integrity is upheld. Conversely, it is also important to evaluate the quality and effectiveness of teaching, which has a tangible impact on the learning outcomes of students. AI-driven performance assessment and monitoring tools have been widely developed for both learners and educators. Since a majority of evaluation material are in textual format, NLP-based models in particular have a major presence in this domain. We divide this section into student-focused and teacher-focused approaches, depending on the direct focus group of such applications.

(i) \textbf{Student-focused}: 

\textit{Knowledge tracing}. An effective way of monitoring the learning progress of students is through knowledge tracing, which models knowledge development in students in order to predict their ability to answer the next problem correctly given their current mastery level of knowledge concepts.
This not only benefits the students by identifying areas they need to work on, but also the educators in designing targeted exercises, personalized learning recommendations and adaptive teaching strategies \cite{liu2019ekt}. An important step of such systems is cognitive modeling, which models the latent characteristics of students based on their current knowledge state. Traditional approaches for cognitive modeling include factor analysis methods which estimate student knowledge by learning a function (logistic in most cases) based on various factors related to the students, course materials, learning and forgetting behavior, etc. \cite{cen2006learning,pavlik2009performance,pavlik2005practice}. Another research direction explores Bayesian inference approaches that update student knowledge states using probabilistic graphical models like Hidden Markov Model (HMM) on past performance records \cite{corbett1994knowledge}, with substantial research being devoted to personalizing such model parameters based on student ability and exercise difficulty \cite{yudelson2013individualized,khajah2014integrating}. Recommender system techniques based on matrix factorization have also been proposed, which predict future scores given a student-exercise performance matrix with known scores \cite{thai2010recommender,toscher2010collaborative}.
\cite{abdelrahman2022knowledge} provides a comprehensive taxonomy of recent work in deep learning approaches for knowledge tracing. Deep knowledge tracing (DKT) was one of the first such models
which used recurrent neural network architectures for modeling the latent knowledge state along with its temporal dynamics to predict future performance \cite{piech2015deep}. Extensions along this direction include incorporating external memory structures to enhance representational power of knowledge states \cite{zhang2017dynamic,abdelrahman2019knowledge}, incorporating attention mechanisms to learn relative importance of past questions in predicting current response \cite{pandey2019self,ghosh2020context}, leveraging textual information from exercise materials to enhance prediction performance \cite{su2018exercise,liu2019ekt} and incorporating forgetting behavior by considering factors related to timing and frequency of past practice opportunities \cite{nagatani2019augmenting,shen2021learning}. Graph neural network based architectures were recently proposed in order to better capture dependencies between knowledge concepts or between questions and their underlying knowledge concepts \cite{nakagawa2019graph,yang2020gikt,tong2020structure}. 
Specific to programming, \cite{wang2017learning} used a sequence of embedded program submissions to train RNNs to predict performance in the current or the next programming exercise.
However as pointed out in \cite{abdelrahman2022knowledge}, handling of non-textual content as in images, mathematical equations or code snippets to learn richer embedding representations of questions or knowledge concepts remains relatively unexplored in the domain of knowledge tracing.

\textit{Grading and feedback}. While technological developments have made it easier to provide content to learners at scale, scoring their submitted work and providing feedback on similar scales remains a difficult problem. While assessing multiple-choice and fill-in-the-blank type questions is easy enough to automate, automating assessment of open-ended questions (e.g. short answers, essays, reports, code samples, etc.) and questions requiring multi-step reasoning (e.g. theorem proving, mathematical derivations, etc.) is equally hard. 
%Moreover, it is expensive and time-consuming to collect data annotated by human experts to train AI systems for such automatic evaluation. 
But automatic evaluation remains an important problem not only because it reduces the burden on teaching assistants and graders, but also removes grader-to-grader variability in assessment and helps accelerate the learning process for students by providing real-time feedback \cite{srikant2014system}.

In the context of written prose, a number of Automatic Essay Scoring (AES) and Automatic Short Answer Grading (ASAG) systems have been developed to reliably evaluate compositions produced by learners in response to a given prompt, and are typically trained on a large set of written samples pre-scored by expert raters \cite{dikli2006overview,shermis2003automated}. 
Over the last decade, AI-based essay grading tools evolved from using handcrafted features such as word/sentence count, mean word/sentence length, n-grams, word error rates, POS tags, grammar, punctuation, etc. \cite{adamson2014automated,phandi2015flexible,cummins2016constrained,contreras2018automated} to automatically extracted features using deep neural network variants \cite{taghipour2016neural,dasgupta2018augmenting,nadeem2019automated,uto2020robust}.
Such systems have been developed not only to provide holistic scoring (assessing essay quality with a single score), but also for more fine-grained evaluation by providing scoring along specific dimensions of essay quality, such as organization \cite{persing2010modeling}, prompt-adherence \cite{persing2014modeling}, thesis clarity \cite{persing2013modeling}, argument strength \cite{persing2015modeling}, thesis strength \cite{ke2019give}, etc.
Since it is often expensive to obtain expert-rated essays to train on each time a new prompt is introduced, considerable attention has been given to cross-prompt scoring using multi-task, domain adaptation or transfer learning techniques, both with handcrafted \cite{phandi2015flexible,cummins2016constrained} and automatically extracted features \cite{li2020sednn,song2020multi}.
Moreover feedback being a critical aspect of essay drafting and revising, AES systems are increasingly being adopted into Automated Writing Evaluation (AWE) systems that provide formative feedback along with (or instead of) final scores and therefore have greater pedagogical usefulness \cite{hockly2019automated}. For example, AWE systems have been developed for providing feedback on errors in grammar, usage and mechanics \cite{burstein2004automated} and text evidence usage in response-to-text student writings \cite{zhang2019erevise}.

AI-based evaluation tools are also heavily used in computer science education, particularly programming, due to its inherent structure and logic. Traditional approaches for automated grading of source codes such as test-case based assessments \cite{douce2005automatic} and assessments using code metrics (lines of code, number of variables, number of statements, etc.), while simple, are neither robust nor effective at evaluating program quality. 

A more useful direction measures similarities between abstract representations (control flow graphs, system dependence graphs) of the student's program and correct implementations of the program \cite{wang2007semantic,vujovsevic2013software} for automatic grading. Such similarity measurements could also be used to construct meaningful clusters of source codes and propagate feedback on student submissions based on the cluster they belong to \cite{huang2013syntactic,mokbel2013domain}. \cite{srikant2014system} extracts informative features from abstract representations of the code to train machine learning models using expert-rated evaluations in order to output a finer-grained evaluation of code quality. % based on a problem-independent rubric. 
\cite{piech2015learning} used RNNs to learn program embeddings that can be used to propagate human comments on student programs to orders of magnitude more submissions. A bottleneck in automatic program evaluation is the availability of labeled code samples. Approaches proposed to overcome this issue include learning question-independent features from code samples \cite{singh2016question,tarcsay2022use} or zero-shot learning using human-in-the-loop rubric sampling \cite{wu2019zero}.

Elsewhere, driven by the maturing of automatic speech recognition technology, AI-based assessment tools have been used for mispronunciation detection in computer-assisted language learning \cite{li2009high,li2016detecting,zhang2020end} or the more complex problem of spontaneous speech evaluation where the student's response is not known apriori \cite{shashidhar2015automatic}. Mathematical language processing (MLP) has been used for automatic assessment of open response mathematical questions \cite{lan2015mathematical,baral2021improving}, mathematical derivations \cite{tanclustering} and geometric theorem proving \cite{mendis2017automatic}, where grades for previously unseen student solutions are predicted (or propagated from expert-provided grades), sometimes along with partial credit assignment. \cite{zhang2022automatic}, moreover, overcomes the limitation of having to train a separate model per question by using multi-task and meta-learning tools that promote generalizability to previously unseen questions.

\textit{Academic integrity issues}. Another aspect of performance assessment and monitoring is to ensure the upholding of academic integrity by detecting plagiarism and other forms of academic or research misconduct. \cite{foltynek2019academic} in their review paper on academic plagiarism detection in text (e.g. essays, reports, research papers, etc.) classifies plagiarism forms according to an increasing order of obfuscation level, from verbatim and near-verbatim copying to translation, paraphrasing, idea-preserving plagiarism and ghostwriting. In a similar fashion, plagiarism detection methods have been developed for increasingly complex types of plagiarism, and widely adopt NLP and ML-based techniques for each \cite{foltynek2019academic}. For example, lexical detection methods use n-grams \cite{alzahrani2015arabic} or vector space models \cite{vani2014using} to create document representations that are subsequently thresholded or clustered \cite{vani2014using} to identify suspicious documents. Syntax-based methods rely on PoS tagging \cite{gupta2014using}, frequency of PoS tags \cite{hurlimann2015glad} or comparison of syntactic trees \cite{tschuggnall2013detecting}. Semantics-based methods employ techniques such as word embeddings \cite{ferrero2017usingword}, Latent Semantic Analysis \cite{soleman2014experiments}, Explicit Semantic Analysis \cite{meuschke2017analyzing}, word alignment \cite{sultan2014dls}, etc., often in conjunction with other ML-based techniques for downstream classification \cite{alfikri2014detailed,hanig2015exb}. Complementary to such textual analysis-based methods, approaches that use non-textual elements like citations, math expressions, figures, etc. also adopt machine learning for plagiarism detection \cite{pertile2016comparing}. \cite{foltynek2019academic} also provides a comprehensive summary of how classical ML algorithms such as tree-based methods, SVMs, neural networks, etc. have been successfully used to combine more than one type of detection method to create the best-performing meta-system. More recently, deep learning models such as different variants of convolutional and recurrent neural network architectures have also been used for plagiarism detection \cite{el2020new,el2022reliable}.

In computer science education where programming assignments are given to evaluate students, source code plagiarism can also been classified based on increasing levels of obfuscation \cite{faidhi1987empirical}. The detection process typically involves transforming the code into a high-dimensional feature representation followed by measurement of code similarity. Aside from tradionally used features extracted based on structural or syntactic properties of programs \cite{ji2007source,lange2007using}, NLP-based approaches such as n-grams \cite{ohmann2015efficient}, topic modeling \cite{ullah2021intelligent}, character and word embeddings \cite{manahi2021deep} and character-level language models \cite{katta2018machine} are increasingly being used for robust code representations. Similarly for downstream similarity modeling or classification, unsupervised \cite{acampora2015fuzzy} and supervised \cite{bandara2011machine,manahi2021deep} machine learning and deep learning algorithms are popularly used.

It is worth noting that AI itself makes plagiarism detection an uphill battle. With the increasing prevalence of easily accessible large language models like InstructGPT \cite{ouyang2022training} and ChatGPT \cite{chatgpt2022} that are capable of producing natural-sounding essays and short answers, and even working code snippets in response to a text prompt, it is now easier than ever for dishonest learners to misuse such systems for authoring assignments, projects, research papers or online exams. How plagiarism detection approaches, along with teaching and evaluation strategies, evolve around such systems remains to be seen.

(ii) \textbf{Teacher-focused}: Teaching Quality Evaluations (TQEs) are important sources of information in determining teaching effectiveness and in ensuring learning objectives are being met. The findings can be used to improve teaching skills through appropriate training and support, and also play a significant role in employment and tenure decisions and the professional growth of teachers. Such evaluations have been traditionally performed by analyzing student evaluations, teacher mutual evaluations, teacher self-evaluations and expert evaluations \cite{hu2021teaching}, which are labor-intensive to analyze at scale. Machine learning and deep learning algorithms can help with teacher evaluation by performing sentiment analysis of student comments on teacher performance \cite{esparza2017sentiment,gutierrez2018mining,onan2020mining}, which provides a snapshot of student attitudes towards teachers and their overall learning experiences. Further, such quantified sentiments and emotional valence scores have been used to predict students' recommendation scores for teachers in order to determine prominent factors that influence student evaluations \cite{okoye2022towards}. \cite{vijayalakshmi2020predicting} uses student ratings related to class planning, presentation, management and student participation to directly predict instructor performance.

Apart from helping extract insights from teacher evaluations, AI can also be used to evaluate teaching strategies on the basis of other data points from the learning process. For example, \cite{duzhin2018machine} used a symbolic regression-based approach to evaluate the impact of assignment structures and collaboration type on student scores, which course instructors can use for the purpose of self-evaluation. Several works use a combination of student ratings and attributes related to the course and the instructor to predict instructor performance and investigate factors affecting learning outcomes \cite{mardikyan2011analyzing,ahmed2016using,abunasser2022prediction} .

\subsubsection{Outcome prediction}

While a course is ongoing, one way to assess knowledge development in students is through graded assignments and projects. On the other hand, educators can also benefit from automatic prediction of students' performance and automatic identification of students at risk of course non-completion. This can be accomplished by monitoring students' patterns of engagement with the course material in association with their demographic information. Such apriori understanding of a student's outcome allows for designing effective intervention strategies. Presently, most K-12, undergraduate and graduate students, when necessary resources are available, rely on computer and web-based infrastructure \cite{bulman2016technology}. A rich source of data indicating student state is therefore generated when a student interacts with the course modules. Prior to computers being such an integral component in education, researchers frequently used surveys and questionnaires to gauge student engagement, sentiment and attrition probability. In this section we will summarize research developments in the field of AI that \textit{generate early prediction of student outcomes - both final performance and possibility of drop-out}. 

Early research in outcome prediction focused on building explanatory regression-based models for understanding student retention using college records \cite{dey1993statistical}. The active research direction in this space gradually shifted to tackling the more complex and more actionable problems of understanding whether a student will complete a program \cite{dekker2009predicting}, estimating the time a student will take to complete a degree \cite{herzog2006estimating} and predicting the final performance of a student \cite{nghe2007comparative} given the current student state. In the subsequent paragraphs, we will be discussing the research contributions for outcome prediction with distinction between performance prediction in assessments and course attrition prediction. Note that we discuss these separately as poor performance in any assessment cannot be generalized into a course non-completion.

(i) \textbf{Apriori performance prediction}: Apriori prediction of performance of a student has several benefits - it allows a student to evaluate their course selection, allows educators to evaluate progress and offer additional assistance as needed etc. Not surprisingly therefore AI-based methods have been proposed to automate this important task in the education process. 

Initial research articles predicting performance estimated time to degree completion \cite{herzog2006estimating} using student demographic, academic, residential and financial aid information, student parent data and school transfer records. In a related theme, researchers have also mapped the question of performance prediction into a final exam grade prediction problem (i.e., excellent, good, fair, fail etc.) \cite{nghe2007comparative, bydvzovska2016comparative, dien2020deep}. This granular prediction eventually allows educators to assess which students require additional tutoring. Baseline algorithms in this context are Decision Trees, Support Vector Machines, Random Forests, Artificial Neural Networks etc (regression or classification based on the problem setup). Researchers have aimed to improve the performance of the predictors by including relevant information such as student engagement, interactions \cite{ramesh2013modeling, bydvzovska2016comparative}, role of external incentives \cite{jiang2014predicting}, previous performance records \cite{tamhane2014predicting} etc. \cite{xu2017progressive} proposed that a student's performance or when the student anticipates graduation should be predicted progressively (using an ensemble machine learning method) over the duration of the student's tenure as the academic state of the student is ever-evolving and can be traced through their student records. The process of generalizing performance prediction to non-traditional modes of learning such as hybrid or blended learning and on-line learning has benefitted from the inclusion of additional information sources such as web-browsing information \cite{trakunphutthirak2019study}, discussion forum activity, student study habits \cite{gitinabard2019widely}etc. 

In addition to exploring a more informative and robust feature set, recently, deep learning based approaches have been identified to outperform traditional machine learning algorithms. For example, \cite{waheed2020predicting} used deep feed-forward neural networks and split the problem of predicting student grade into multiple binary classification problems viz., Pass-Fail, Distinction-Pass, Distinction-Fail, Withdrawn-Pass. \cite{tsiakmaki2020transfer} analyzed if transfer learning (i.e., pre-training neural networks on student data on a different course) can be used to accurately predict student performance. \cite{chui2020predicting} used a generative adversarial network based architecture, ICGAN-DSVM, to address the challenges of low volume of training data in alternative learning paradigms such as supportive learning. \cite{dien2020deep} proposed extensive data pre-processing using min-max scaler, quantile transformation etc. before passing the data in a deep-learning model such as one-dimensional convolutional network (CN1D) or recurrent neural networks (LSTM). For a comprehensive survey of ML approaches for this topic, we would refer readers to  \cite{rastrollo2020analyzing} and \cite{hellas2018predicting}. 

(ii) \textbf{Apriori attrition prediction}: Students dropping out before course completion is a concerning trend. This is more so in developing nations where very few students finish primary school \cite{knofczynski2017dropout}. The outbreak of the COVID-19 pandemic exacerbated the scenario due to indefinite school closures. This led to loss in learning and progress towards providing access to quality education \cite{moscoviz2022learning}. The causes for dropping out of a course or a degree program can be diverse, but early prediction of it allows administrative staff and educators to intervene. To this end, there have been efforts in using machine learning algorithms to predict attrition. 

\textit{Massive Open Online Courses (MOOCs)}: In the context of attrition, special mention must be made of Massive Open Online Courses (MOOCs). While MOOCs promise the democratization of education, one of the biggest concerns with MOOCs is the disparity between the number of students who sign up for a course versus the number of students who actually complete the course - the drop-out rate in MOOCs is significantly high \cite{hollands2018benefits}, \cite{reich2019mooc}. Yet in order to make post-secondary and professional education more accessible, MOOCs have become more a practical necessity than an experiment. The COVID-19 pandemic has only emphasized this necessity \cite{purkayastha2021unstoppable}. In our literature search phase, we found a sizeable number of contributions in attrition prediction that uses data from MOOC platforms. In this subsection, we will be including those as well as attrition prediction in traditional learning environments. 

Early educational data mining methods \cite{dekker2009predicting} proposed to predict student drop-out mostly used data sources such as student records (i.e., student demographics, academic, residential, gap year, financial aid information etc.) and administrative records (major administrative changes in education, records of student transfers etc.) to train simple classifiers such as Logistic Regression, Decision Tree, BayesNet, Random Forest etc. Selecting an appropriate set of features and designing explainable models has been important as these later inform intervention\cite{aguiar2015and}. To this end, researchers have explored features such as students’ prior experiences, motivation and home environment \cite{deboer2013bringing}, student engagement with the course \cite{aguiar2014engagement}, \cite{ramesh2014learning} etc. In an exciting experiment, \cite{veeramachaneni2014towards} crowd-sourced feature engineering for predicting student attrition. With the inclusion of an online learning component (particularly relevant for MOOCs), click-stream data and browser information generated allowed researchers to better understand student behavior in an ongoing course. Using historical click-stream data in conjuction with present click-stream data, allowed \cite{kloft2014predicting} to effectively predict drop-outs weekly using a simple Support Vector Machine algorithm. This kind of data has also been helpful in understanding the traits indicative of decreased engagement \cite{sinha2014capturing}, the role of a social cohort structure \cite{yang2013turn} and the sentiment in the student discussion boards and communities \cite{wen2014sentiment} leading up to student drop-out. \cite{he2015identifying} addresses the concern that weekly prediction of probability of a student dropping out might have wide variance by including smoothing techniques. On the other hand, as resources to intervene might be limited, \cite{lakkaraju2015machine} recommends assigning a risk-score per student rather than a binary label. \cite{brooks2015time} considers the level of activity of a student in bins of time during a semester as a binary features (active vs. inactive) and then uses these sequences as n-grams to predict drop-out. Recent developments in predicting student attrition propose the use of data acquired from disparate sources in addition to more sophisticated algorithms such as deep feed-forward neural networks \cite{imran2019predicting}, hybrid logit leaf model \cite{coussement2020predicting} etc.

\section{Impact of COVID-19 pandemic on driving AI research in the frontier of education}

COVID-19 pandemic, possibly the most significant social disruptor in recent history, impacted more than 1.5 billion students worldwide \cite{unesco2022unescos} and is believed to have had far-reaching consequences in the domain of education, possibly even generational setbacks \cite{tadesse2020impact, dorn2021covid, spector2022new}. As lockdowns and social distancing mandated a hastened transition to fully virtual delivery of educational content, teachers and administrators grappled with  the issue of providing quality education at scale in new formats while ensuring learning progress remains unhindered. We discuss how the pandemic changed the education landscape through remote learning, exacerbated socio-economic inequalities in learning and outline how AI has the potential to address some of the identified issues. 

\textbf{Remote learning and AI}. The pandemic era saw an increasing adoption of video conferencing softwares and social media platforms for knowledge delivery, combined with more asynchronous formats of learning. These alternative media of communication were often accompanied by decreasing levels of engagement and satisfaction of learners \cite{wester2021student,hollister2022engagement}. There was also a corresponding decrease in practical sessions, labs and workshops, which are quite critical in some fields of education. For example, \cite{hilburg2020medical} outlines the disruptive effect of the pandemic on the medical education landscape, possibly making medical students under-prepared for clinical training experiences typically conducted in person, such as history-taking, physical examination, etc.

However, the pandemic also led to an accelerated adoption of AI-based approaches in education. Pilot studies show that the pandemic led to a significant increase in the usage of AI-based e-learning platforms \cite{pantelimon2021evolution}. Moreover, a natural by-product of the transition to online learning environments is the generation and logging of more data points from the learning process \cite{xie2020covid} that can be used in AI-based methods to assess and drive student engagement and provide personalized feedback. Online teaching platforms also make it easier to incorporate web-based content, smart interactive elements and asynchronous review sessions to keep students more engaged \cite{pantelimon2021evolution,kexin2020future}.

\textbf{Inclusive education and AI}. Several recent works have investigated the role of pandemic-driven remote and hybrid instruction in widening gaps in educational achievements by race, poverty level and gender \cite{halloran2021pandemic,goldhaber2022consequences,unesco2021when}. A widespread transition to remote learning necessitates access to proper infrastructure (electricity, internet connectivity and smart electronic devices that can support video conferencing apps and basic file sharing) as well as resources (learning material, textbooks, educational softwares, etc.), which create barriers for low-income groups \cite{munoz2021remote}. \cite{munoz2021remote} also outlines the importance of parents as allies of educators to mitigate some of the limitations of remote learning. This is far less achievable in low-income groups (and ethnic minorities in some cases) due to a lack of awareness, lower educational qualifications of adults in the household, as well as disproportionate loss of income and public health impact of the COVID-19 pandemic in these communities \cite{luengo2021artificial}. Even within similar populations,  unequal distribution of household chores, income-generating activities and access to technology-enabled devices affect students of different genders disproportionately \cite{unesco2021when}. Moreover, remote learning requires a level of tech-savviness on the part of students and teachers alike, which might be less prevalent in people with learning disabilities. It is therefore undeniable that such widespread changes had far-reaching impacts on inclusive and quality education worldwide. 

In this context, \cite{garg2020impact} outlines the different ways AI is used in special need education for development of adaptive and inclusive pedagogies. \cite{salas2022artificial} reviews the different ways in which AI positively impacts education of minority students, e.g. through facilitating performance/engagement improvement, student retention, student interest in STEM/STEAM fields, etc. \cite{salas2022artificial} also outlines the technological, pedagogical and socio-cultural barriers for AIEd in inclusive education.

\section{Discussion}

In this article, we have investigated the involvement of artificial intelligence in the end-to-end educational process. We have highlighted specific research problems both in the planning and in the knowledge delivery phase and reviewed the technological progress in addressing those problems in the past two decades. To the best of our knowledge, such distinction between proactive and reactive phases of education accompanied by a technical deep-dive is an uniqueness of this review. 

\subsection{Major trends in involvement of AI in the end-to-end education process}
The growing interest in AIEd can be inferred from Fig. \ref{count_by_year} and Fig. \ref{review_articles_vs_year} which show how both the count of technical contributions and the count of review articles on the topic have increased over the past two decades. It is to be noted that the number of technical contributions in 2021 and 2022 (assuming our sample of reviewed articles is representative of the population) might have fallen in part due to pandemic-related indefinite school closures and shift to alternate learning models. This triggered a setback on data collection, reporting and annotation efforts due to a number of factors including lack of direct access to participants, unreliable network connectivity, necessity of enumerators adopting to new training modes, etc \cite{wolf2022remote}. 
Another important observation from Fig. \ref{across_target_review} is that AIEd research in most categories focuses heavily on learners in universities, e-learning platforms and MOOCs - work targeting pre-school and K-12 learners is conspicuously absent. A notable exception is research surrounding tutoring aids that has a nearly uniform attention for different target audience groups.

In all categories, to different extents, we see a distinct shift from rule-based and statistical approaches to classical ML to deep learning methods, and from handcrafted features to automatically extracted features. This advancement goes hand-in-hand with the increasingly complex nature of the data being utilized for training AIEd systems. Whereas earlier approaches used mostly static data (e.g. student records, administrative records, demographic information, surveys and questionnaires), the use of more sophisticated algorithms necessitated (and in turn benefited from) more real-time and high-volume data (e.g. student-teacher/peer-peer interaction data, click-stream information, web-browsing data, etc.). 
The type of data used by AIEd systems also evolved from mostly tabular records to more text-based and even multi-modal data, spurred on by the emergence of large language models that can handle large quantities of such data. 

Even though data-hungry models like deep neural networks have grown in popularity across almost all categories discussed here, AIEd often suffers from the availability of sufficient labeled data to train such systems. This is particularly true for small classes and new course offerings, or when existing curriculum or tests are changed to incorporate new elements. As a result, another emerging trend in AIEd focuses on using information from resource-rich courses or existing teaching/evaluation content through domain adaptation, transfer learning, few-shot learning, meta learning, etc.

\subsection{Existing challenges in adopting Artificial Intelligence for education} 

In 2023, artificial intelligence has permeated the lives of people in some aspect or other globally (e.g. chat-bots for customer service, automated credit score analysis, personalized recommendations etc.). At the same time, AI-driven technology for the education sector is gradually becoming a practical necessity globally. The question therefore is, what are the existing barriers in global adoption of AI for education in a safe and inclusive manner, such that SDG4 can be achieved \cite{united2015goal}. In this section, we analyze and discuss some of our observations with regards to existing AIEd technologies and the challenges in deploying these technologies globally. 

\textbf{Lack of cultural awareness in AIEd tools}: The lack of international representation in the AIEd research field has been brought to light several times through both perspective review articles \cite{blanchard2012weird, blanchard2015socio} as well as through bibliometric analysis \cite{chen2022two, prahani2022artificial}. In particular, most of the existing AIEd technologies were conceptualized and developed in the context of Western, Educated, Industrialized, Rich and Democratic (WEIRD) societies. Therefore, such technologies make implicit assumptions about students and teachers (e.g., individualism, fairness, cooperation, sensitivity to disciplinary actions etc.) that are valid in such demographics. Moreover, it is likely that the training data for the AI entities in these frameworks was obtained from a very specific section of the global population. As a result, the trained models will have limited generalization capabilities when deployed for a different target audience.  Classroom instruction strategies, teacher-student interactions, peer to peer interactions, teaching curriculum are expected to vary widely across the globe making cultural awareness and sensitivity in educational frameworks a must \cite{blanchard2010infusing, ogan2015preface, mohammed2019towards}. Beyond classroom and online courses, culture aware tutoring systems are relevant for professional and vocational training as well. The notion of developing culturally aware educational technologies is relatively new and some early research in this direction \cite{mohammed2008using, young2011significance, si2015virtual} proposes the embodiment of culturally sensitive gestures and movements in intelligent tutoring systems. However, it remains to be seen how this concept matures and is made available in classroom and online settings for learners from different backgrounds \cite{shum2019learning}.

\textbf{Lack of concrete legal and ethical guidelines for AIEd research}: As pointed out by \cite{pedro2019artificial}, besides most AIEd researchers being concentrated in the technologically advanced parts of the world, most AIEd platforms and applications are owned currently by the private sector. The private investor funded research in big corporations such as Coursera, EdX, IBM, McGraw-Hill and start-ups like Elsa, Century, Querium have yielded several robust AIEd applications. However, as these platforms are privately owned, there is little transparency and regulations regarding their development and operations. Due to this, there is growing concern on the part of guardians and teaching staff regarding the data accessed by these platforms, privacy and security of the data stored and explainability of the deployed models. To alleviate this, regulation policies at the international, national and state levels can help address the concerns of the end users. While many tech-savvy nations have had a head start in this \cite{stirling2017government}, drafting general guidelines for AIEd platforms is still very much a nascent concept for most policy makers. 

\textbf{Lack of equitable access to infrastructure hosting AIEd}: Education is one of the most important social equalizers \cite{winthrop2018leapfrogging}. However, in order to ensure more people have access to quality education, AI-enabled teaching and studying tools are necessary to reduce the stress on educators and administrative staff \cite{pedro2019artificial}. The irony here is that the cost of deploying and operating AIEd tools often alienates communities with limited means thereby widening the gap in access to education. \cite{nye2015intelligent} mentions that access to electricity, internet, data storage and processing hardware have been barriers in deploying AI-driven platforms. To remove these obstacles, changes must be brought about in local and global levels. While formation of international alliances that invest in infrastructure development can usher in the technology in developing nations, changes in local policies can expedite the process \cite{mbangula2022adopting}. 

\textbf{Lack of skilled personnels to operate AIEd tools in production}: Investing in AIEd research and supporting infrastructure alone is not sufficient to ensure long term utility and usage of AI-driven tools for education. Workforce responsible for using these tools on a day-to-day basis must also be brought up to speed. Currently, there is a considerable amount of apprehension, particularly in developing countries, regarding use of AI for education \cite{shum2019learning, alam2021possibilities}. The main concerns are related to data privacy and security, job security, ethics etc. post adoption of AI in this sector. These concerns in turn have slowed down integration of technology for education. In this context, we must echo \cite{pedro2019artificial} in mentioning that while these concerns are relevant and must be addressed, in our review of AIEd research, we have not found any evidence that should invoke panic in educators and administrative staff. AIEd research as it stands today only augments the role of the teacher, and does not eliminate it. Furthermore, for the foreseeable future, we would need a human in the loop to provide feedback and ensure proper daily usage of these tools. 

\subsection{Concluding Remarks}

Through this review, we identified the paradigm shift over the past 20 years in formulating computational models (i.e., choice of algorithms, choice of features etc.) and training them (i.e., choice of data) - we are indeed increasingly leaning towards \textit{sophisticated yet explainable }frameworks. As the scope of this review includes a period of social disruption due to COVID-19 pandemic, it provided us the opportunity to introspect on the utility and the robustness of the proposed technology thus far. To this end, we have discussed the concerns and limitations brought to light by the pandemic and research ideas spawning from that. 

With the target of ensuring equitable access to education being set for 2030, one of the inevitable questions arising is: \textit{are we ready to use AI driven ed-tech tools to support educators and students?}. This remains however a question to be answered. Based on our survey, we have observed that while in some parts of the world we have seen great momentum in making AIEd a part and parcel of the education sector, in other parts of the world this progress is stymied by inadequate access to necessary infrastructure and human resources. The pivotal point at this time is that while there needs to be changes at a socio-economic level to adopt the state of the art AI driven ed-tech technologies as standard tools for education, the progress made and the ongoing conversations are reasons for optimism.

%Bibliography
\bibliographystyle{unsrt}

\end{document}